\documentclass[prd,twocolumn,floatfix,preprintnumbers,showpacs]{revtex4}
\pdfoutput=1
\usepackage{graphicx}
\usepackage{dcolumn}
\usepackage{bm}
\usepackage{color}
\usepackage{epsfig}
\usepackage{hyperref}
\usepackage{amsmath} 

\def\lsim{\mathrel{\mathop
  {\hbox{\lower0.5ex\hbox{$\sim$}\kern-0.8em\lower-0.7ex\hbox{$<$}}}}}
\def\gsim{\mathrel{\mathop
  {\hbox{\lower0.5ex\hbox{$\sim$}\kern-0.8em\lower-0.7ex\hbox{$>$}}}}}

\def\lapprox{\hbox{\lower .8ex\hbox{$\,\buildrel < \over\sim\,$}}}
\def\gapprox{\hbox{\lower .8ex\hbox{$\,\buildrel > \over\sim\,$}}}

\def\xisp{\xi(\sigma, \pi)}

\def\xips{\xi(\pi,\sigma)}

\newcommand{\nc}{\newcommand}

\nc{\be}[1]{\begin{equation}\mbox{$\label{#1}$}}
\nc{\bea}[1]{\begin{eqnarray} \mbox{$\label{#1}$}}
\nc{\Section}[2]{\section{#2}\label{#1}}
\nc{\Bibitem}[1]{\bibitem{#1}}
\nc{\Label}[1]{\label{#1}}

\nc{\Mpc}{Mpc/h}
\nc{\vev}[1]{\langle #1 \rangle}

\nc{\eea}{\end{eqnarray}}
\nc{\ee}{\end{equation}}

\begin{document}


\title{First Cosmological Constraints on Dark Energy \\ 
       from the Radial Baryon Acoustic Scale}

\author{Enrique Gazta\~naga}
\email{gazta@ice.cat}
\affiliation{Institut de Ci\`encies de l'Espai (IEEC-CSIC),
Facultat de Ci\`encies, Torre C5, Campus UAB, E-08193 Bellaterra (Barcelona), Spain}
\author{Ramon Miquel}
\email{rmiquel@ifae.es}
\affiliation{Instituci\'o Catalana de Recerca i Estudis Avan\c{c}ats,
Institut de F\'{\i}sica d'Altes Energies,
Edifici Cn, Facultat de Ci\`encies, Campus UAB, 
E-08193 Bellaterra (Barcelona), Spain}
\author{Eusebio S\'anchez}
\email{Eusebio.Sanchez@ciemat.es}
\affiliation{Centro de Investigaciones Energ\'eticas, Medioambientales y Tecnol\'ogicas,
Avenida Complutense 22, E-28040 Madrid, Spain}

\date{\today}

\pacs{95.36.+x, 98.65.Dx, 98.80.Es}
\begin{abstract}
We present cosmological constraints arising from the first
measurement of the  radial (line-of-sight)  baryon acoustic 
oscillations (BAO) scale in the large scale structure traced
by the galaxy distribution.
Here we use these radial BAO measurements at $z=0.24$ and
$z=0.43$ to derive new
constraints on dark energy and its equation of state for a 
flat universe, without any other assumptions on the cosmological model:
$w = -1.14 \pm 0.39$ (assumed constant), $\Omega_m = 0.24^{+0.06}_{-0.05}$.
If we drop the assumption of flatness and include
previous cosmic microwave background and supernova data, 
we find $w = -0.974 \pm 0.058$, $\Omega_m = 0.271 \pm 0.015$, and
$\Omega_k = -0.002 \pm 0.006$, in good agreement with a flat cold dark matter
cosmology with a cosmological consant.
To our knowledge, these are the most stringent 
constraints on these parameters to date under our stated assumptions.
\end{abstract}

\maketitle


Before recombination, the Universe was filled by a plasma of coupled photons and
baryons. On horizon crossing, cosmological fluctuations produced sound
waves in this plasma and when recombination occurred, 380\,000 years after the Big Bang,
the distance covered by the sound wave was about 150 comoving Mpc. This is the so called
sound horizon $r_s$, also known as the BAO scale $r_{BAO}$. This signature can be found
both in the cosmic microwave background (CMB) and large scale structure (LSS), 
so the baryon acoustic peak can be used as a standard 
ruler in the Universe. The value of $r_{BAO}$ depends on a few
physical quantities, mostly the time to recombination and the time
of matter domination, both of which are well known from measurements of
CMB temperature anisotropies, i.e.\ through parameters $\Omega_m h^2$ and $\Omega_B h^2$,
also in good agreement with large scale structure and primordial abundance measurements.
Thus the spectrum of CMB fluctuations  provide an accurate estimate of
$r_{BAO}= 153 \pm 1.9$ Mpc, independent of the value of $H_0$, dark energy equation
of state $w$ or the curvature of the Universe \cite{wmap5}. 
At the same time the CMB also provides an angular
measurement of the BAO scale, which together with $r_{BAO}$, can be used for
geometrical test, such as the measurement of curvature.

A series of recent papers \cite{paper1,paper2,paper3,paper4} have presented
the clustering of Luminous Red Galaxies (LRG, see \cite{eisenstein2001})
in the latest spectroscopic Sloan Digital Sky Survey (SDSS) data releases, 
DR6 \& DR7, which include over 75\,000 LRG 
galaxies and sample over 1 ${\,\rm Gpc^3/h^3}$
to z=0.47. 
The last paper~\cite{paper4} focuses on the study of the 2-point correlation function $\xisp$,
separated in perpendicular $\sigma$ and line-of-sight $\pi$ directions to
find a significant detection of a peak at $r\simeq 110$~Mpc/h 
(for $H_0=100$~h~Km/s),
which shows as a circular ring in the $\sigma$-$\pi$ plane.
There is also a significant detection of the peak along the line-of-sight (radial) 
direction both in sub-samples at low, $z$=0.15-0.30, and high redshifts, 
$z=0.40-0.47$.
The overall shape and location of the peak in the 2-point and 3-point function
are consistent with its originating from the recombination-epoch baryon acoustic oscillations.
This has been used to produce, for the first time, a direct measurement
of the Hubble parameter $H(z)$ as a function of redshift.  This is based on
a calibration to the BAO scale measured in the CMB \cite{wmap5}.
The values of $H(z)$ are then used in \cite{paper4} to compare to other cosmological
constraints that provide estimates for $H_0$.

Here, instead of calibrating the BAO distance, we will use the direct 
dimensionless
measurement as a redshift scale (shown in Table \ref{tab:rbao}) to provide
cosmological constraints that are independent of $H_0$. This is somehow similar
to what is done in \citet{percival} for the spherically-averaged 
(i.e.\ monopole)
BAO detection.  Both the 2-Degree-Field Galaxy Redshift Survey
2dFGRS and SDSS spectroscopic redshift surveys have been used
to constrain cosmological parameters at BAO distances
\cite{detection,clusteringlrg,sanchez,hutsia, hutsib}. 
Photometric LRG catalogs have also been used to obtain cosmological constraints 
\cite{padmanabhan2007,blake}.

None of these earlier papers have constrained the baryon feature in the 
line-of-sight (LOS) direction.
We will argue here that the radial BAO (hereafter RBAO) detection provides in practice fully independent 
constraints from the monopole (or angular) BAO measurements. It is therefore
important to check if such independent constrains agree with the flat concordance
cosmological constant model ($\Lambda$CDM) and, if so, what are the new combined constraints
when we add the new data to previous results. Also note that in previous BAO
detections the constraints come from either a fit to the overall
shape of the correlation function or to the normalized wiggles in the
power spectrum $P(k)$,
while in the LOS measurement of $r_{BAO}$ in~\cite{paper4} they
only used the BAO peak location (identified
by the position of maximum signal to noise around $90-120$ Mpc/h) and 
not the shape. There are potential advantages and disadvantages of both methods.
Using a fit to the overall shape is a more stringent test of the model, but
puts more weight on the intermediate
scales, where the statistical errors are smaller but where the modeling is more
uncertain. Fitting the
peak location is more model independent but, precisely because of this, provides
little evidence for the physical origin of such peak. In the case
of the analysis in \cite{paper4}, the physical evidence comes 
from  the agreement between model and data 
both on the radial direction (see Figs.~13--15 in \cite{paper4})
and on the overall $\xips$ plane, which shows a clear
BAO ring (with the right monopole contribution) that blends into the 
extended LOS peak (see Fig.~10 in \cite{paper4}). A model with no
radial BAO is disfavored at $3.2\sigma$.


\vspace{1em}

To work with fiducial comoving scales, \cite{paper4} uses a reference flat 
$\Lambda$CDM Hubble rate:
\begin{equation}
H_{ref}(z) = H_{0}\sqrt{\Omega_m (1+z)^3+1-\Omega_m}
\end{equation}
with $\Omega_m = 0.25$ to determine fiducial values for $r_{BAO}$ at two 
redshifts. 
Here, $\Delta z_{BAO}(z)$ are obtained from $r_{BAO}(z)$ in~\cite{paper4} as
$\Delta z_{BAO}(z) = r_{BAO}(z) H_{ref}(z)/c$. Note that
the measurements of $\Delta z_{BAO}(z)$ do not depend on $H_0$, since the
dependences contained in $r_{BAO}(z)$ and $H_{ref}(z)/c$ cancel out.

Previous BAO (e.g.\ \cite{detection,percival}) analyses have 
only looked at the monopole, which is
the average of $\xi(\sigma, \pi)$ over orientations:
\begin{equation}
\xi_0 (r) = \int_{0}^1 \xi(\sigma, \pi) d\mu
\end{equation}
where $r = \sqrt{\sigma^2 + \pi^2}$ and $\mu = \pi/r$.
The BAO peak found by \cite{paper4} was located at 
radial $\pi \simeq 110$~Mpc/h for  $\sigma<5$~Mpc/h, which
corresponds to $\mu>0.999$.  The contribution of radial
modes with $\mu>0.999$
to the monopole at $r=110$ in the above integral is 
less than $1\%$ even if we take the amplitude of $\xisp$
to be 10 times larger in the radial $\pi$ direction than
in the transverse $\sigma$ direction. Moreover, the covariance
between radial and monopole correlations has been 
shown to be less that 10$\%$ on scales larger than 40 Mpc/h
(see Fig.~2 of \cite{Sanchez09}).
Thus, in practice,
we can regard here the BAO measurements from the monopole as
independent from the RBAO measurements.


\vspace{1em}

Once the radial BAO scale has been measured~\cite{paper4}, there are 
several possible approaches to extract the cosmological parameters solely from 
this measurement. Percival {\it et al.}~\cite{percival} propose three 
possible ways. We have used two of them, that are described below. The
third one uses the ratio of BAO scales at different redshifts. We have 
not used it, since we have the BAO scale at two different 
redshifts, so only one ratio can be constructed, and the degeneracy 
in the determination of cosmological parameters is larger using this approach.

\begin{table}[ht]
\begin{center}
 \begin{tabular}{|c|c||c|c|c||c|c|c|}
\hline
 Sample  & $z_m$ & $r_{BAO}$ & $\sigma_{st}$  & $\sigma_{sys}$ &                  &                &                \\
 z range &       & Mpc/h    &                &               & $\Delta z_{BAO}$  &  $\sigma_{st}$  & $\sigma_{sys}$ \\
\hline
\hline
 0.15-0.30 & 0.24 & 110.3 & 2.9 & 1.8 & 0.0407 & 0.0011 & 0.0007\\
 0.40-0.47 & 0.43 & 108.9 & 3.9 & 2.1 & 0.0442 & 0.0015 & 0.0009\\
\hline
\end{tabular}
\caption{The BAO fiducial scale $r_{BAO}$ in the LOS direction calculated 
with a flat reference $H_{ref}(z)$ cosmology
of $\Omega_m=0.25$, for two redshift slices: $z_m$ is the respective pair-weighted mean redshift,
and $\sigma_{st}$ and $\sigma_{sys}$ are the statistical and systematic errors on $r_{BAO}$ (from 
\cite{paper4}). Here we use 
the direct $\Delta z_{BAO}$ measurement, shown in the sixth column, which relates to 
the fiducial scale as $\Delta z_{BAO} = r_{BAO} H_{ref}(z)/c$
and is independent of the value chosen for $H_{ref}(z)$. 
The last two columns show the corresponding errors.
\label{tab:rbao}
}
\end{center}
\end{table}

We will derive cosmological constraints from the measured values of
$\Delta z_{BAO}$ shown in Table~\ref{tab:rbao}, which can be expressed
as 
\begin{equation}
\Delta z_{BAO}(z_i) = \frac{H(z_i) r_s}{c} \ ,
\label{eq:deltaz1} 
\end{equation}
where $r_s$ is the sound horizon at recombination. 
It is important to remark that the two 
measurements of $\Delta z_{BAO}$ are independent, due to the analysis 
strategy used in~\cite{paper4}, which determines the BAO scale in two
well separate intervals of redshift.

\begin{table}[!t]
\begin{center}
\begin{tabular}{|l||c|c||c|c||c|c|}
\hline
Data Set      & $\Omega_m$ & $\sigma(\Omega_m)$ & $\Omega_k$   & $\sigma(\Omega_k)$ & $w$          & $\sigma(w)$        \\
\hline
\hline
\begin{tabular}{l}
RBAO \\
a) 
\end{tabular} & 0.242      &$^{+0.061}_{-0.053}$  & $0$  & ---                & $-1.14$     &$^{+0.38}_{-0.40}$     \\ 
\hline
\begin{tabular}{l}
 RBAO \\
b) 
\end{tabular} & 0.248      &$\pm 0.053$        & $0$  & ---                & $-1.16$     &$^{+0.42}_{-0.45}$     \\ 
\hline
\begin{tabular}{l}
RBAO + \\
WMAP5~\cite{wmap5}
\end{tabular} & 0.274      &$^{+0.035}_{-0.036}$  & $0$  & ---                & $-0.92$    &$^{+0.16}_{-0.22}$     \\ 
\hline
\begin{tabular}{l}
RBAO + \\
WMAP5 + \\
SNe~\cite{kowalski08}
\end{tabular} & 0.268      &$^{+0.015}_{-0.014}$  & $0$  & ---                & $-0.961$    &$^{+0.056}_{-0.058}$   \\
\hline
\hline
\begin{tabular}{l}
RBAO \\
a) \\
\end{tabular} & 0.249      &$^{+0.049}_{-0.036}$  & $-0.08$      &$^{+0.25}_{-0.18}$    & $-1$ &  ---               \\ 
\hline
\begin{tabular}{l}
RBAO \\
b) 
\end{tabular} & 0.223      &   $\pm 0.099$      & $-0.028$    &$\pm 0.058$          & $-1$ &  ---               \\ 
\hline
\begin{tabular}{l}
RBAO + \\
WMAP5~\cite{wmap5}
\end{tabular} & 0.264      &$^{+0.017}_{-0.016}$   &$-0.0025$    &$\pm 0.0059$         & $-1$ &  ---               \\ 
\hline
\begin{tabular}{l}
RBAO + \\
WMAP5 + \\
SNe~\cite{kowalski08}
\end{tabular} & 0.271      &$^{+0.015}_{-0.014}$   &$-0.0026$    &$\pm 0.0060$         & $-1$ &  ---              \\
\hline
\hline
\begin{tabular}{l}
RBAO + \\
WMAP5~\cite{wmap5}
\end{tabular} & 0.244      &$^{+0.062}_{-0.052}$  &$-0.0049$     &$^{+0.0121}_{-0.0061}$   & $-1.13$  & $^{+0.37}_{-0.39}$   \\ 
\hline
\begin{tabular}{l}
RBAO + \\
WMAP5 + \\
SNe~\cite{kowalski08}
\end{tabular} & 0.271      &$^{+0.015}_{-0.014}$  &$-0.0021$     & $\pm 0.0062 $      & $-0.974$  & $^{+0.057}_{-0.059}$ \\
\hline
\end{tabular}
\caption[]{Results from the cosmology fits with the radial BAO (RBAO) data and additional
data sets assuming a constant equation of state parameter $w$. In the data set a) the BAO radial data are
used, and the values and errors of $\Omega_mh^2$ and $\Omega_bh^2$ from~\cite{wmap5} 
are input into the computation of $H(z_i)r_s$. 
For the data set b) the BAO radial data is combined with the 
measurement of $l_A(z^*)$ from~\cite{wmap5}. In all cases, the WMAP5 measurements within the appropriate
cosmology model (flat, cosmological constant, etc.) are used, although the differences are minute. See text
for details. The other rows combine the RBAO redshift scales
presented in Table~\ref{tab:rbao} with
external data sets. All fits have values of $\chi^2/N_{dof}$ close to one. In particular, the fit with RBAO, WMAP5 and SNe data
within a non-flat constant-$w$ cosmology (last row) has $\chi^2/N_{dof} = 312/306$. 
\label{tab:cosmo}
}
\end{center}
\end{table}
%
\begin{figure}[!t]
\begin{center}
\includegraphics[width=7.0truecm]{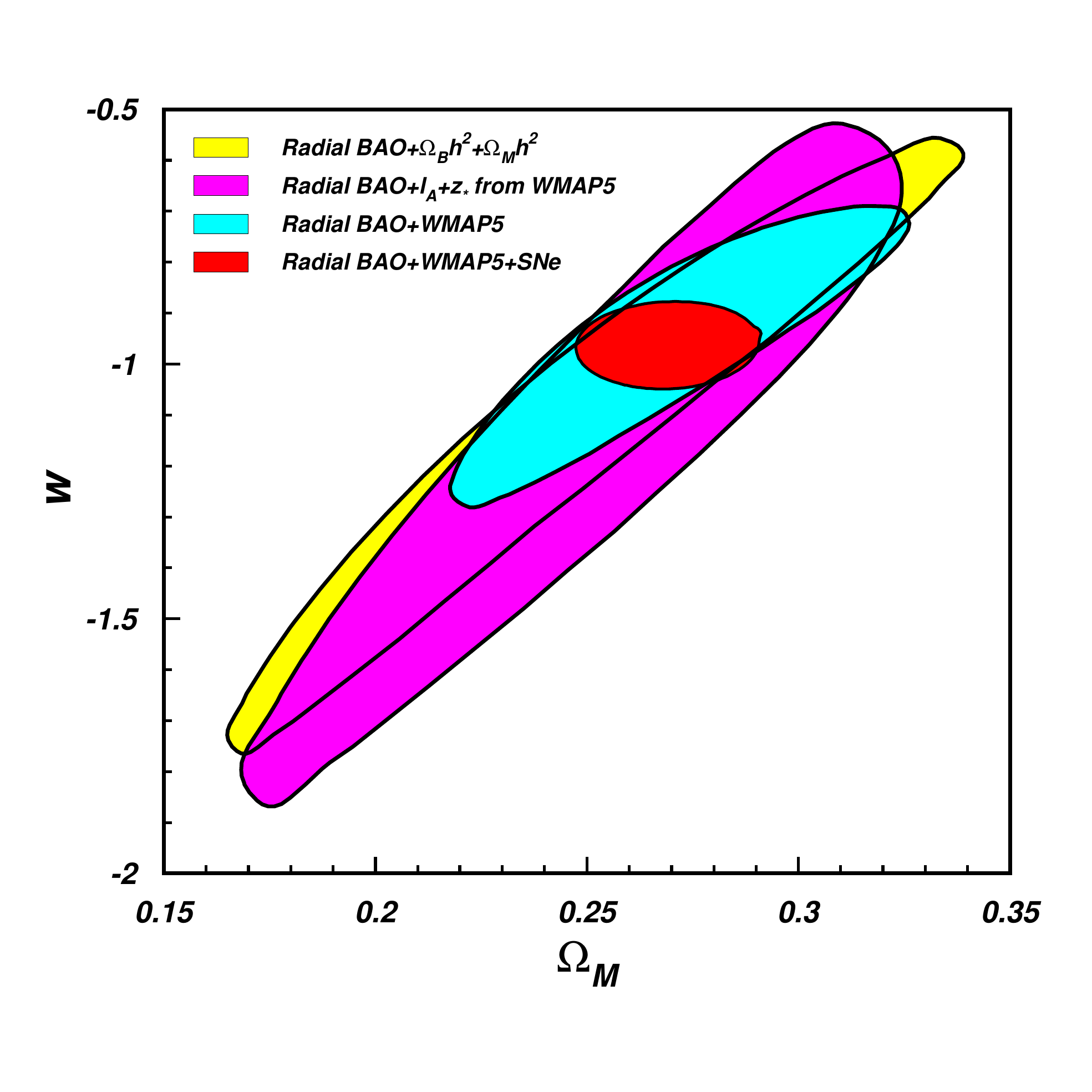}
\end{center}
\caption{68\% CL constraints on the plane $\Omega_m$--$w$ when a flat
Universe is assumed. In yellow, from the measurements of the RBAO scale plus
$\Omega_bh^2$ and $\Omega_mh^2$. In pink, from RBAO plus $l_A$ and $z^*$
from WMAP5~\cite{wmap5}
In blue, from the RBAO scale plus WMAP5 and
in red from the RBAO scale, plus WMAP5 plus supernovae~\cite{kowalski08}. 
Results are consistent
among them and compatible with a $\Lambda$CDM cosmology.}
\label{fig:flat_wcdm}
\end{figure}
Throughout this letter, we will assume a standard 
Friedmann-Lema\^{\i}tre-Robertson-Walker cosmology having as free
parameters the matter density $\Omega_m$, curvature $\Omega_k$, Hubble constant
$h$ (with $H_0 = 100\,h$~km/s/Mpc), and the parameter $w$ of the equation of 
state of dark energy, i.e.~the
ratio of its pressure to its density $w = p_{DE}/\rho_{DE}$. 
Unless otherwise specified,
a constant $w$ is assumed, and thus, the standard $\Lambda$CDM model is recovered
for $w=-1$. We will start by assuming also a flat Universe, $\Omega_k = 0$. 


In order to constrain 
the dark energy parameters $\Omega_\Lambda(\equiv 1-\Omega_m)$ and $w$, we can 
take two approaches:
\begin{itemize}
\item[a)] We can express $r_s$ in (\ref{eq:deltaz1}) 
as~\footnote{We have checked that
we can safely neglect the terms proportional
$\Omega_k$ and $\Omega_{DE}$ in Eq.~\ref{eq:rs}.}
\begin{equation}
r_s = \frac{c}{(3\Omega_mH_0^2)^{1/2}}\int_0^{\frac{1}{1+z_d}}\frac{da}
{(a+a_{eq})^{1/2}(1+a\frac{3\Omega_bh^2}{4\Omega_\gamma h^2})^{1/2}}
\label{eq:rs}
\end{equation}
with
\begin{equation}
a_{eq} = \frac{\Omega_\gamma h^2(1+0.2271 N_{eff})}{\Omega_mh^2}
\end{equation}
with $z_d$, the so-called drag redshift, written in terms of $\Omega_bh^2$ and $\Omega_mh^2$ (Eq.~(3)
in~\cite{wmap5}), and take $\Omega_bh^2 = 0.02273 \pm 0.0066$ and 
$\Omega_mh^2=0.1329 \pm 0.0064$ from 
the five-year results of the Wilkinson Microwave Anisotropy Probe (WMAP5)~\cite{wmap5} as external constraints
(we could have also used determinations from nucleosynthesis or LSS measurements),
and fix $\Omega_\gamma h^2 = 2.449\times 10^{-5}$ and $N_{eff} = 3.04$. 
Note that the expression for $\Delta z_{BAO}(z_i)$ in~Eq.~(\ref{eq:deltaz1}) is independent of
$H_0$. Once the one-sigma WMAP5 constraints on $\Omega_bh^2$ and $\Omega_mh^2$ are included in the fit,
the only remaining unknowns are $\Omega_\Lambda(\equiv 1-\Omega_m)$ and $w$. Since
we have two independent $\Delta z_{BAO}(z_i)$, we will be able to determine them.
\item[b)] Alternatively, we can use as external constraints 
the measurements by WMAP5 of the ratio $l_A(z^*)$ between the distance to the last scattering
surface and $r_s(z^*)$ and of $z^*$ itself,
$l_A (z^*) = 302.14 \pm 0.87$ and $z^* = 1090.5 \pm 1.0$ with a $\sim 40$~\% positive correlation, when assuming a flat CDM model with 
constant equation of state $w$.
Again, $H_0$
cancels out and we are left with only $\Omega_m$ and $w$ as 
unknowns, which we then proceed to determine. 
\end{itemize}

In all cases, a frequentist $\chi^2$ fit including correlations
is applied to the data, using the
publicly available Minuit code \cite{minuit} for minimization and contour calculation. Systematic
errors in $\Delta z_{BAO}(z_i)$ have been neglected,
being subdominant (see Table~\ref{tab:rbao}) and relatively uncertain. 
Including them does not change the results significantly.
The results 
obtained for $\Omega_m$ and $w$ and their 1-$\sigma$ errors, 
$\sigma(\Omega_m)$ and $\sigma(w)$, are shown in rows 1 and 2 of Table~\ref{tab:cosmo}. 
Both determinations, a) and b), are 
consistent between them and in agreement with $\Lambda$CDM. The two-dimensional
constraints at 68\% CL can be seen in Fig.~\ref{fig:flat_wcdm}.
\begin{figure}[!tbh]
\centering
\includegraphics[width=7.0truecm]{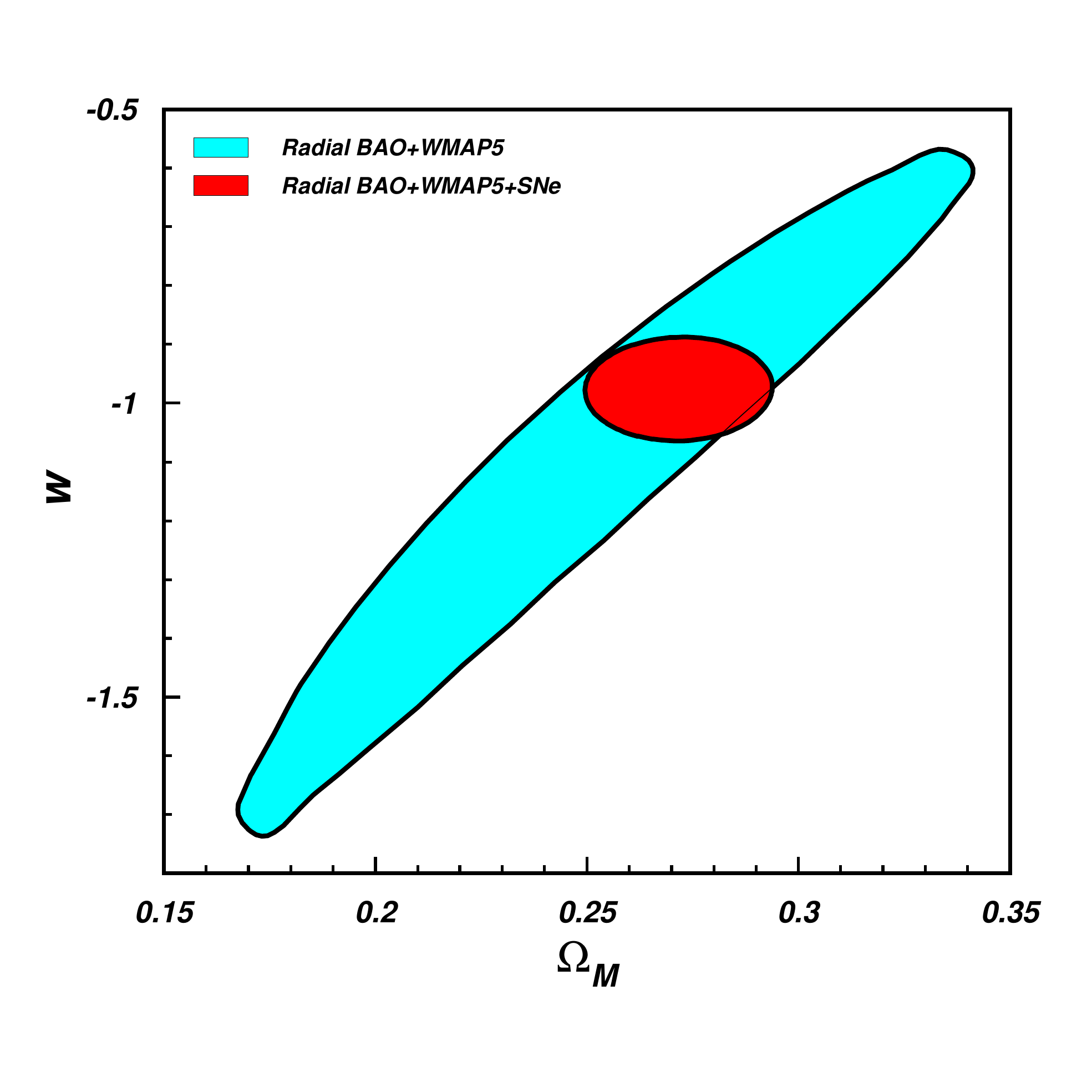}
\caption[]{68\% CL constraints on the plane $\Omega_m$--$w$ when a non-flat
constant-$w$ cosmology is assumed. 
In blue, from the RBAO scale plus WMAP5~\cite{wmap5} and
in red from the RBAO scale, plus WMAP5 plus supernovae~\cite{kowalski08}. 
Results are consistent
with a cosmological constant Universe with a precision around 5\%.}
\label{fig:nonflatocdm1}
\end{figure}
%
%
%

\vspace{1em}

Next, we have combined the results obtained from the radial BAO scale with the full CMB distance
constraints, using the WMAP5 measurements of the shift parameter $R(z^*)$, 
the acoustic scale $l_A(z^*)$ and the redshift at decoupling $z^*$
presented in Table~10 of~\cite{wmap5}, and the covariance matrix in Table~11
therein.
We have also
added the supernovae set compiled by Kowalski {\it et al.}~\cite{kowalski08}, which can be
found in~\footnote{{\tt http://supernova.lbl.gov/Union/}}.
We have taken their data covariance matrix without systematical errors. Adding systematics does not
change the results qualitatively.
The results
of the combination assuming flatness can be seen in rows 3 and 4 of
Table~\ref{tab:cosmo}. We determine the equation of state parameter $w$,
assumed constant, to be consistent with a cosmological constant to about 5\%.

If, instead, we assume the $\Lambda$CDM model and drop the assumption of
flatness, we obtain the results in rows 5--8 of Table~\ref{tab:cosmo}, which show
consistency with a flat Universe at the 0.5\% level.

We can now keep the equation of state parameter $w$ free and
drop the assumption of flatness. The constraints, shown in the last two rows of Table~\ref{tab:cosmo}
remain very robust when combining the three data sets.
The two-dimensional $\Omega_m$--$w$ contour
at 68\% CL is shown in 
Fig.~\ref{fig:nonflatocdm1}. 
All results are in good agreement with $\Lambda$CDM. 
Moreover, the combination allows
us to simultaneously 
determine $w$ with a precision around 5\%, and set the flatness of the Universe
with a precision of about half of a percent. 
As a by-product, we obtain a Hubble parameter
$h = 0.703 \pm 0.020$, consistent with most recent results in the literature.
If instead of RBAO we use in the combination the BAO
monopole determination in~\cite{detection}, we find compatible results
with errors that are 20--40\% larger (see Table~6 in~\cite{kowalski08}).
On the other hand, if we combine RBAO and the monopole in~\cite{detection},
we find no improvement in the error band with respect to the case with RBAO
alone.

Finally, we can drop the assumption of constant $w$ and, instead, use the 
parametrization $w(z) = w_0 + w_a\cdot z/(1+z)$ \cite{chevallier01,linder03}. 
Assuming a flat
Universe, and combining RBAO with WMAP5 and supernova data, we get 
$w_0 = -1.05^{+0.11}_{-0.10}$ and $w_a = 0.44^{+0.37}_{-0.51}$, with an 
87\% anticorrelation, consistent with $\Lambda$CDM ($w_0 = -1$, $w_a = 0$). We conclude that
current data show no evidence for a time-dependent $w$, although the constraint on $w_a$
is still very weak.
%
%
\vspace{1em}

In summary, we have determined for the first time the cosmological parameters $\Omega_m$ and
$w$ using the radial BAO scale, to be $\Omega_m = 0.24^{+0.06}_{-0.05}$ and 
$w = -1.14 \pm 0.39$. These results are perfectly consistent with the
expectations of the $\Lambda$CDM cosmology.
Moreover, when these results are combined with the cosmological distance constraints coming from
the CMB~\cite{wmap5} and type-Ia supernovae~\cite{kowalski08}, we measure
$\Omega_m = 0.271 \pm 0.015$, $\Omega_k = -0.002 \pm 0.006$, and
$w = -0.974\pm 0.058$. The result is in good agreement
with a flat $\Lambda$CDM cosmology,
and shows that the determination of the RBAO scale is consistent
with all the other cosmological measurements. 

In the future, currently planned BAO surveys like Baryon Oscillations Sloan Survey 
(BOSS)~\footnote{{\tt http://www.sdss.org/news/releases/20080110.sdss3.html}}
or Physics of the Accelerating Universe (PAU)~\cite{pau} will measure radial BAOs with
greater precision up to redshifts close to $z=1$, producing more stringent constraints on the properties of
dark energy.
%
%
\vspace{1em}

We would like to thank the ``Centro de Ciencias de Benasque 
Pedro Pascual'' for the hospitality while most of this paper was written.
RM wants to thank Eric Linder for useful discussions.
We acknowledge funding from the Spanish Ministerio de Ciencia
y Tecnolog\'{\i}a (MEC) through projects AYA2006-06341, AYA2006-13735-C02-01 and
AYA2006-13735-C02-02, and Consolider-Ingenio CSD2007-00060, together with
EC-FEDER funding, and the research project 2005SGR00728
from Generalitat de Catalunya. 
\bibliography{tesi}
\end{document}